\documentclass[preprint,prd]{revtex4-1}

\usepackage{xcolor}
\usepackage{graphicx}
\usepackage{multirow}

\usepackage{array}
\newcolumntype{L}{>{\displaystyle}l}
\newcolumntype{C}{>{\displaystyle}c}
\newcolumntype{R}{>{\displaystyle}r}

\newcommand{\comma}{{\quad , \quad}}

\renewcommand{\d}[1]{\mathinner{d#1}}
\newcommand{\fn}[2]{\mathinner{#1\mathopen{\left(#2\right)}}}
\newcommand{\eq}[1]{Eq.~(\ref{#1})}
\newcommand{\eqs}[2]{Eqs.~(\ref{#1}) and (\ref{#2})}
\newcommand{\eqss}[3]{Eqs.~(\ref{#1}), (\ref{#2}) and (\ref{#3})}

\newcommand{\RX}{\mathcal{R}_{\delta\rho_X}}
\newcommand{\RY}{\mathcal{R}_{\delta\rho_Y}}
\newcommand{\Rm}{\mathcal{R}_{\delta\rho_\mathrm{m}}}
\newcommand{\Rr}{\mathcal{R}_{\delta\rho_\mathrm{r}}}
\newcommand{\Rt}{\mathcal{R}_{\delta\rho}}

\newcommand{\Smr}{\mathcal{S}_\mathrm{mr}}

\newcommand{\pc}{\mathinner{\mathrm{pc}}}
\newcommand{\kpc}{\mathinner{\mathrm{kpc}}}
\newcommand{\Mpc}{\mathinner{\mathrm{Mpc}}}

\newcommand{\eV}{\mathinner{\mathrm{eV}}}

\newcommand{\GeV}{\mathinner{\mathrm{GeV}}}
\newcommand{\mpl}{M_\mathrm{Pl}}

\begin{document}

\title{Effects of thermal inflation on small scale density perturbations}
\author{Sungwook E. Hong}
\affiliation{School of Physics, Korea Institute for Advanced Study\\
Seoul 130-722, Republic of Korea}
\author{Hyung-Joo Lee}
\author{Young Jae Lee}
\affiliation{Department of Physics, KAIST\\
Daejeon 305-701, Republic of Korea}
\author{Ewan D. Stewart}
\affiliation{Department of Physics\\
University of Auckland\\
Auckland 1010, New Zealand\\
on sabbatical leave from\\
Department of Physics, KAIST\\
Daejeon 305-701, Republic of Korea}
\author{Heeseung Zoe}
\email{heezoe@dgist.ac.kr}
\affiliation{School of Basic Science\\
Daegu Gyeongbuk Institute of Science and Technology (DGIST)\\
Daegu 711-873, Republic of Korea}
\affiliation{Division of General Studies\\
Ulsan National Institute of Science and Technology (UNIST)\\
Ulsan 689-798, Republic of Korea}

\date{\today}

\begin{abstract}
In cosmological scenarios with thermal inflation, extra eras of moduli matter domination, thermal inflation and flaton matter domination exist between primordial inflation and the radiation domination of Big Bang nucleosynthesis.
During these eras, cosmological perturbations on small scales can enter and re-exit the horizon, modifying the power spectrum on those scales.
The largest modified scale, $k_\mathrm{b}$, touches the horizon size when the expansion changes from deflation to inflation at the transition from moduli domination to thermal inflation.
We analytically calculate the evolution of perturbations from moduli domination through thermal inflation and evaluate the curvature perturbation on the constant radiation density hypersurface at the end of thermal inflation to determine the late time curvature perturbation.
Our resulting transfer function suppresses the power spectrum by a factor $\sim 50$ at $k \gg  k_\mathrm{b}$, with $k_\mathrm{b}$ corresponding to anywhere from megaparsec to subparsec scales depending on the parameters of thermal inflation.
Thus, thermal inflation might be constrained or detected by small scale observations such as CMB distortions or 21cm hydrogen line observations.
\end{abstract}

\pacs{98.80.Bp, 98.80.Cq}

\maketitle

\section{Introduction}

Thermal inflation \cite{Lyth:1995hj,Lyth:1995ka,Yamamoto:1985mb,Yamamoto:1985rd,Enqvist:1985kz,Bertolami:1987xb,Ellis:1986nn,Ellis:1989ii,Randall:1994fr}, a brief low energy inflation that occurs when thermal effects hold an unstable flat direction at the origin, occurs naturally in supersymmetric theories and is motivated to solve the moduli and gravitino problems \cite{Coughlan:1983ci,Banks:1993en,de Carlos:1993jw}.
Additionally, thermal inflation provides a mechanism for baryogenesis \cite{Stewart:1996ai,Jeong:2004hy,Felder:2007iz,Kim:2008yu,Kawasaki:2006py,Lazarides:1985ja,Yamamoto:1986jw,Mohapatra:1986dg} and has implications for dark matter \cite{Yamamoto:1985mb,Yamamoto:1986jw,Kim:2008yu}.
When thermal inflation is included, observable cosmology starts as usual with primordial inflation, which generates the density perturbations that go on to form galaxies, etc., but then has additional eras of moduli matter domination, thermal inflation and flaton matter domination inserted before the usual radiation domination of Big Bang nucleosynthesis (BBN).

During these thermal inflation eras, cosmological perturbations on small scales can enter and re-exit the horizon, modifying the power spectrum on those scales, while perturbations on larger scales, corresponding to cosmic microwave background (CMB) or large scale structure (LSS) observations, remain outside the horizon, preserving and only modestly redshifting their spectrum.

The observational impact of thermal inflation on the gravitational wave background was studied in \cite{Mendes:1998gr,Easther:2008sx}.
Thermal inflation wipes out any potentially observable gravitational waves from primordial inflation on solar system or smaller scales \cite{Mendes:1998gr}, but the first order phase transition at the end of thermal inflation generates gravitational waves \cite{Kosowsky:1991ua,Easther:2008sx} with frequencies in the Hz range, though their amplitude may be too small for them to be observed in the near future \cite{Easther:2008sx}.

In this paper, we study the impact of thermal inflation on small scale density perturbations with the aim of finding signatures of, or constraints on, thermal inflation.
We note that a variety of small scale physics, such as ultracompact minihalos or primordial black holes \cite{Carr:1975qj,Josan:2009qn,Bringmann:2011ut}, lensing dispersion of SNIa \cite{Ben-Dayan:2013eza}, CMB distortions \cite{Chluba:2011hw,Chluba:2012gq} and the 21cm hydrogen line at or prior to the era of reionization \cite{Cooray:2006km, Mao:2008ug} could be used as tools for studying the effects of thermal inflation on the small scale power spectrum.
Several prominent upcoming observations are designed for such small scale physics, for example the Primordial Inflation Explorer (PIXIE) \cite{Kogut:2011xw} and the Polarized Radiation Imaging and Spectroscopy Mission (PRISM) \cite{Andre:2013afa, Andre:2013nfa} for CMB distortions and the Square Kilometre Array (SKA) \cite{Furlanetto:2009qk} for the 21cm hydrogen line.

In Section~\ref{sec:history}, we outline cosmology with thermal inflation and determine its characteristic scales.
In Section~\ref{sec:background}, we analytically calculate the thermal inflation transfer function which modifies the primordial power spectrum.
In Section~\ref{sec:discussion}, we summarize our results and briefly discuss the possibilities of observing the effects of thermal inflation.
In this paper, we set $\hbar = c = 8 \pi G = \mpl = 1$.

\section{Cosmology with thermal inflation}
\label{sec:history}

\begin{figure}
\centering
\includegraphics[width=0.8\textwidth]{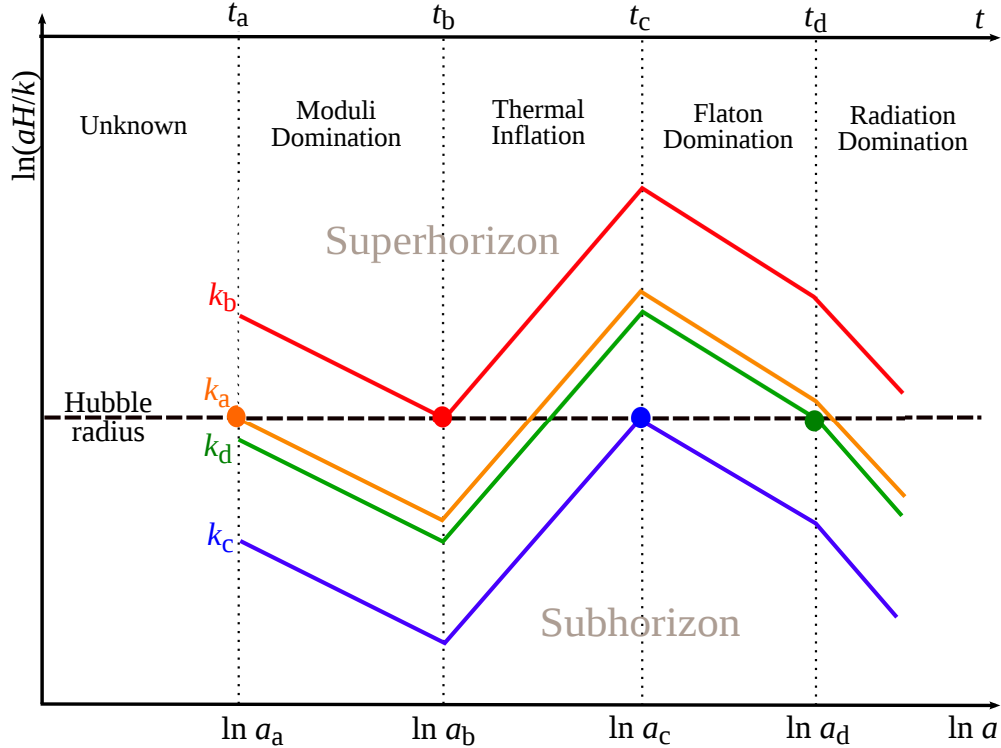}
\caption{ \label{fig:4scales}
Characteristic scales of cosmology with thermal inflation.
There are four characteristic scales, $k_\mathrm{a}$, $k_\mathrm{b}$, $k_\mathrm{c}$ and $k_\mathrm{d}$, corresponding to the comoving scale of the horizon at each of the era boundaries.
$k_\mathrm{b}$ is the largest, and hence most observationally relevant, scale, see \eq{kblength}.
}
\end{figure}

In cosmology with thermal inflation \cite{Lyth:1995hj,Lyth:1995ka,Stewart:1996ai,Jeong:2004hy,Felder:2007iz,Easther:2008sx,Kim:2008yu}, there are two different epochs of inflation.
The first, primordial inflation \cite{gliner1,gliner2,Guth:1980zm,Linde:1981mu,Albrecht:1982wi}, generates the primordial perturbations which are observed in the CMB and grow to form galaxies and the LSS.
The second, thermal inflation, is brief and occurs after primordial inflation but before Big Bang nycleosynthesis, at a sufficiently low energy scale to solve the moduli problem, and may affect the perturbations on very small scales as we discuss in the following sections.

In Fig.~\ref{fig:4scales}, we illustrate the thermal inflation eras of moduli domination, thermal inflation and flaton domination.
They are preceded by primordial inflation plus a possible post-inflationary era and followed by the radiation domination of BBN.

For definiteness, we consider a general class of supersymmetry breaking scenarios in which supersymmetry is broken in a hidden sector at a scale $M_\mathrm{s}$ and transmitted to the observable sector via gravitational strength interactions, so that the supersymmetry breaking scale in the observable sector is $m_\mathrm{s} \sim M_\mathrm{s}^2 / \mpl$.
We set $m_\mathrm{s} \sim 10^3 \GeV$.

In the early universe, the finite energy density, represented by the Hubble parameter $H$, breaks supersymmetry.
When $H \gtrsim m_\mathrm{s}$, the supersymmetry breaking by the finite energy density dominates over the vacuum supersymmetry breaking giving a moduli potential of the form
\begin{equation} \label{cosmodpot}
\fn{V}{\Phi} \sim H^2 \mpl^2 \fn{f}{\frac{\Phi}{\mpl}}
\sim H^2 \left( \Phi - \Phi_1 \right)^2 + \cdots .
\end{equation}
However, as the Hubble parameter drops through $H \sim m_\Phi$, the moduli potential switches to its vacuum form
\begin{equation} \label{modpot}
\fn{V}{\Phi} \sim M_\mathrm{s}^4 \fn{g}{\frac{\Phi}{\mpl}}
\sim m_\Phi^2 \left( \Phi - \Phi_2 \right)^2 + \cdots
\end{equation}
with
\begin{equation} \label{mmod}
m_\Phi \sim \frac{M^2_\mathrm{s}}{\mpl}
\sim m_\mathrm{s} \sim 10^3 \GeV
\end{equation}
and $\Phi_1 - \Phi_2 \sim \mpl$.
Thus, at $H \sim m_\Phi$, or $t \sim t_\mathrm{a}$ in the notation of Fig.~\ref{fig:4scales}, the moduli start oscillating with Planckian amplitude, dominating the energy density of the universe.
Furthermore, the moduli have relatively low masses and very weak interactions so this moduli oscillation may be sufficiently long-lived to have disastrous effects on, for example, BBN.
During the era of moduli domination, the moduli abandunce is
\begin{equation}
\frac{n_\Phi}{s} \sim \sqrt{ \frac{\mpl}{m_\Phi}} \sim 10^8 ,
\end{equation}
where $n_\Phi$ is the moduli number density and $s$ is the entropy density,
but not to spoil BBN it should be \cite{Kawasaki:2004qu}
\begin{equation} \label{resolution}
\frac{n_\Phi}{s} \lesssim 10^{-15} \text{ to } 10^{-12} .
\end{equation}
This is called the moduli problem \cite{Coughlan:1983ci,Banks:1993en,de Carlos:1993jw}.

Thermal inflation, which is motivated to solve the moduli problem, is realized by an (almost) flat direction with negative mass-squared at the origin, called a flaton.
Like the Standard Model Higgs field, the flaton potential near the origin is
\begin{equation} \label{flaton}
\fn{V}{\phi} = V_0 - \frac{1}{2} m^2_\phi \phi^2 + \cdots .
\end{equation}
Unlike the Standard Model Higgs field, but like many scalar field directions in the Minimal Supersymmetric Standard Model, the flaton does not have a stabilizing $\phi^4$ term.
Instead, higher order terms, or the renormalisation group running of the flaton mass squared, stabilize the flaton potential at a large field value $\phi_0 \gg m_\phi$.
To have zero energy density at the minimum, we require $V_0 \sim m_\phi^2 \phi_0^2$, and so, for $m_\phi \ll \phi_0 \lesssim \mpl$, we have
\begin{equation}
m_\phi \ll V_0^\frac{1}{4}
\lesssim \sqrt{m_\phi \mpl} \, .
\end{equation}
Taking
\begin{equation}
m_\phi \sim m_\mathrm{s} \sim 10^3 \GeV
\end{equation}
gives
\begin{equation} \label{exptV}
10^3 \GeV \ll V_0^\frac{1}{4}
\lesssim 10^{11} \GeV .
\end{equation}
At the finite temperature of the early universe, the flaton's potential is modified by its (assumed) unsuppressed couplings to the thermal bath
\begin{equation} \label{flatoneffective}
\fn{V}{\phi,T} = V_0 + \frac{1}{2} \left( \sigma^2 T^2 - m_\phi^2 \right) \phi^2 + \cdots ,
\end{equation}
where $\sigma$ is not small.
When $T \gtrsim m_\phi/\sigma$, corresponding to $t \lesssim t_\mathrm{c}$ in Fig.~\ref{fig:4scales}, the flaton is trapped at $\phi = 0$, leaving the potential energy $V = V_0$ which drives the thermal inflation.

At $t \sim t_\mathrm{a}$, the moduli dominate the universe, but with a comparable amount of radiation
\begin{equation}
\fn{\rho_\mathrm{m}}{t_\mathrm{a}}
\sim \fn{\rho_\mathrm{r}}{t_\mathrm{a}}
\gg V_0 \, .
\end{equation}
As the universe expands, the moduli and radiation are diluted
\begin{equation}
\rho_\mathrm{m} \propto a^{-3}
{\quad , \quad}
\rho_\mathrm{r} \propto a^{-4} .
\end{equation}

At $t \sim t_\mathrm{b}$, the moduli density drops below $V_0$
\begin{equation}
V_0 \sim  \fn{\rho_\mathrm{m}}{t_\mathrm{b}}
\gg \fn{\rho_\mathrm{r}}{t_\mathrm{b}}
\end{equation}
and thermal inflation begins.
As the universe inflates, the temperature drops
\begin{equation}
T \propto a^{-1}
\end{equation}
and the moduli are diluted to a safely small abundance, see \eq{bbmod} below.

At $t \sim t_\mathrm{c}$, the temperature drops to the critical temperature
\begin{equation} \label{Tc}
T_\mathrm{c} \sim \frac{m_\phi}{\sigma} ,
\end{equation}
at which a first order phase transition ends thermal inflation \cite{Easther:2008sx} and the flaton rapidly rolls towards and oscillates about its minimum at $\phi = \phi_0$, leading to a flaton matter dominated era
\begin{equation}
\rho_\phi \propto a^{-3} .
\end{equation}
The change in state at the end of thermal inflation perturbs the moduli potential by an amount
\begin{equation}
\fn{\delta V}{\Phi}
\sim V_0 \fn{h}{\frac{\Phi}{\mpl}}
\sim \frac{V_0}{\mpl} \Phi + \cdots
\end{equation}
regenerating a moduli abundance
\begin{equation}
\frac{n_\Phi}{n_\phi} \sim \frac{V_0 m_\phi}{m_\Phi^3 \mpl^2} .
\end{equation}

At $t \sim t_\mathrm{d}$, the flaton decays to radiation at a temperature \cite{Kim:2008yu}
\begin{equation}
T_\mathrm{d} \sim 10^2 \text{ to } 10^{-2} \GeV
\end{equation}
and the standard cosmic history of radiation domination, BBN, etc., follows.

The moduli generated at $t \sim t_\mathrm{a}$ are diluted by thermal inflation to an abundance \cite{Lyth:1995ka}
\begin{eqnarray}
\frac{n_\Phi}{s} & \sim & \frac{\fn{g_{*s}}{T_\mathrm{c}} T_\mathrm{c}^3 T_\mathrm{d} \mpl^{1/2}}{V_0 m_\Phi^{1/2}}
\\
& \sim & 10^{-9} \left( \frac{\fn{g_{*s}}{T_\mathrm{c}}}{10^2} \right) \left( \frac{T_\mathrm{c}}{10^3\GeV} \right)^3 \left( \frac{T_\mathrm{d}}{\GeV} \right) \left( \frac{10^3\GeV}{m_\Phi} \right)^\frac{1}{2} \left( \frac{10^7\GeV}{V_0^{1/4}} \right)^4 ,
\end{eqnarray}
where $\fn{g_{*s}}{T}$ is the effective number of entropic degrees of freedom at temperature $T$ \cite{Kolb:1990vq}.
In models of thermal inflation incorporating a mechanism for baryogenesis \cite{Stewart:1996ai,Jeong:2004hy,Felder:2007iz,Kim:2008yu}, entropy release by Affleck-Dine fields during thermal inflation may typically lead to an effective double thermal inflation \cite{Felder:2007iz,Kim:2008yu}, further diluting the moduli by a factor
\begin{eqnarray}
\Delta_\mathrm{AD} & \sim & \left( \frac{\fn{\rho_\mathrm{r}}{T_\mathrm{c}}}{\rho_\mathrm{AD}} \right)^\frac{3}{4}
\sim \left( \frac{\fn{g_*}{T_\mathrm{c}} T_\mathrm{c}^4}{m_{LH_u}^2 \langle LH_u \rangle_\mathrm{TI}} \right)^\frac{3}{4}
\sim \left( \frac{\fn{g_*}{T_\mathrm{c}} T_\mathrm{c}^4 m_\nu}{m_{LH_u}^3 \langle H_u \rangle_\mathrm{EW}^2} \right)^\frac{3}{4}
\\
& \sim & 10^{-8} \left( \frac{\fn{g_*}{T_\mathrm{c}}}{10^2} \right)^\frac{3}{4} \left( \frac{T_\mathrm{c}}{10^3\GeV} \right)^3 \left( \frac{10^3\GeV}{m_{LH_u}} \right)^\frac{9}{4} \left( \frac{174\GeV}{\langle H_u \rangle_\mathrm{EW}} \right)^\frac{3}{2} \left( \frac{m_\nu}{10^{-2}\eV} \right)^\frac{3}{4}
\end{eqnarray}
to
\begin{equation} \label{bbmod}
\frac{n_\Phi}{s} \sim 10^{-17} .
\end{equation}
The moduli regenerated at $t \sim t_\mathrm{c}$ have abundance \cite{Lyth:1995ka}
\begin{equation}
\frac{n_\Phi}{s} \sim \frac{V_0 T_\mathrm{d}}{m_\Phi^3 \mpl^2}
\sim 10^{-18} \left( \frac{T_\mathrm{d}}{\GeV} \right) \left( \frac{10^3\GeV}{m_\Phi} \right)^3 \left( \frac{V_0^{1/4}}{10^7\GeV} \right)^4 .
\end{equation}
Thus for values of $V^{1/4}$ in the middle of its expected range, \eq{exptV}, the moduli can be diluted to, and regenerated with, a sufficiently small abundance to satisfy \eq{resolution}, solving the moduli problem.

\subsection{Characteristic scales}
\label{subsec:characteristic}

The thermal inflation eras, illustrated in Fig.~\ref{fig:4scales}, determine four characteristic scales, $k_\mathrm{a}$, $k_\mathrm{b}$, $k_\mathrm{c}$ and $k_\mathrm{d}$,
where
\begin{equation} \label{kx}
k_x \equiv a_x H_x
\end{equation}
corresponds to the comoving scale of the horizon at the era boundary $t_x$.
We express the $k_x$ in terms of the effective thermal inflation parameters
\begin{equation}
N_{xy} \equiv \ln \frac{a_y}{a_x} \, ,
\end{equation}
which measure the durations of the thermal inflation eras by their number of $e$-folds of expansion, and in turn estimate the $N_{xy}$ in terms of the more fundamental thermal inflation parameters $m_\Phi$, $V_0$, $T_\mathrm{c}$, $T_\mathrm{d}$, etc.

\subsubsection{$k_\mathrm{a}$}

During moduli domination, $t_\mathrm{a} < t < t_\mathrm{b}$, the energy density $\rho \propto a^{-3}$, so
\begin{equation} \label{ka}
k_\mathrm{a} = \frac{a_\mathrm{a}H_\mathrm{a}}{a_\mathrm{b}H_\mathrm{b}} k_\mathrm{b}
= \left( \frac{a_\mathrm{a}}{a_\mathrm{b}} \right)^{-\frac{1}{2}} k_\mathrm{b}
= e^{\frac{1}{2}N_\mathrm{ab}} k_\mathrm{b} \, .
\end{equation}
Using $\rho_\mathrm{a} \sim m_\Phi^2 \mpl^2$ and $\rho_\mathrm{b} \sim V_0$,
\begin{eqnarray}
N_\mathrm{ab} & = & \ln \frac{a_\mathrm{b}}{a_\mathrm{a}}
= \frac{1}{3} \ln \frac{\rho_\mathrm{a}}{\rho_\mathrm{b}}
\simeq \frac{1}{3} \ln \frac{m_\Phi^2 \mpl^2}{V_0}
\\ \label{nab_val}
& \simeq & 11 + \frac{2}{3} \ln \left[ \frac{m_\Phi}{10^3\GeV}
\left( \frac{10^7\GeV}{V_0^{1/4}} \right)^2 \right] .
\end{eqnarray}

\subsubsection{$k_\mathrm{b}$}

During thermal inflation, $t_\mathrm{b} < t < t_\mathrm{c}$, the energy density $\rho = V_0$, so
\begin{equation} \label{kb}
k_\mathrm{b} = e^{-N_\mathrm{bc}} k_\mathrm{c} \, .
\end{equation}
At the beginning of thermal inflation, $t = t_\mathrm{b}$,
\begin{eqnarray}
\rho_\mathrm{r} & = & \frac{\pi^2}{30} \fn{g_*}{T_\mathrm{b}} T_\mathrm{b}^4
\\
& \sim & \left( \frac{V_0}{m_\Phi^2 \mpl^2} \right)^\frac{1}{3} \rho_\mathrm{m} \sim \left( \frac{V_0^2}{m_\Phi \mpl} \right)^\frac{2}{3} ,
\end{eqnarray}
therefore
\begin{eqnarray}
N_\mathrm{bc} & = & \ln \frac{a_\mathrm{c}}{a_\mathrm{b}}
= \ln \frac{T_\mathrm{b}}{T_\mathrm{c}}
\simeq \frac{1}{6} \ln \frac{V_0^2}{\fn{g_*^{3/2}}{T_\mathrm{b}} T_\mathrm{c}^6 m_\Phi \mpl}
\\
& \simeq & 5 + \frac{1}{3} \ln \left[ \left( \frac{10^2}{\fn{g_*}{T_\mathrm{b}}} \right)^\frac{3}{4} \left( \frac{10^3\GeV}{T_\mathrm{c}} \right)^3 \left( \frac{10^3\GeV}{m_\Phi} \right)^\frac{1}{2} \left( \frac{V_0^{1/4}}{10^7\GeV} \right)^4 \right] .
\end{eqnarray}
As discussed above, entropy release during thermal inflation may typically add an extra $\sim 6$ $e$-folds, giving
\begin{equation}
N_\mathrm{bc} \sim 11 .
\end{equation}
More general forms of double or multiple thermal inflation can in principle give even larger values of $N_\mathrm{bc}$.

\subsubsection{$k_\mathrm{c}$}

During flaton domination, $t_\mathrm{c} < t < t_\mathrm{d}$, the energy density $\rho \propto a^{-3}$, so
\begin{equation} \label{kc}
k_\mathrm{c} = \frac{a_\mathrm{c}H_\mathrm{c}}{a_\mathrm{d}H_\mathrm{d}} k_\mathrm{d}
= \left( \frac{a_\mathrm{c}}{a_\mathrm{d}} \right)^{-\frac{1}{2}} k_\mathrm{d}
= e^{\frac{1}{2}N_\mathrm{cd}} k_\mathrm{d} \, .
\end{equation}
Using $\rho_\mathrm{c} = V_0$ and $ \rho_\mathrm{d} = \fn{\rho_\mathrm{r}}{T_\mathrm{d}}$,
\begin{eqnarray} \label{Ncd}
N_\mathrm{cd} & = & \ln \frac{a_\mathrm{d}}{a_\mathrm{c}}
= \frac{1}{3} \ln \frac{\rho_\mathrm{c}}{\rho_\mathrm{d}}
= \frac{1}{3} \ln \frac{30 V_0}{\pi^2 \fn{g_*}{T_\mathrm{d}} T_\mathrm{d}^4}
\\ \label{ncd_val}
& \simeq & 20 + \frac{1}{3} \ln \left[ \frac{10^2}{\fn{g_*}{T_\mathrm{d}}} \left( \frac{\GeV}{T_\mathrm{d}} \right)^4 \left( \frac{V_0^{1/4}}{10^7\GeV}  \right)^4 \right] .
\end{eqnarray}

\subsubsection{$k_\mathrm{d}$}

At the beginning of radiation domination, $t = t_\mathrm{d}$, the scale factor is \cite{Kolb:1990vq}
\begin{equation} \label{ad}
a_\mathrm{d} = \frac{\fn{g_{*s}^{1/3}}{T_0}T_0}{\fn{g_{*s}^{1/3}}{T_\mathrm{d}}T_\mathrm{d}} a_0 \, ,
\end{equation}
where $a_0$ and $T_0$ are the current scale factor and temperature, respectively, and the Hubble parameter is
\begin{eqnarray} \label{Hd}
H_\mathrm{d} = \sqrt{\frac{\rho_\mathrm{d}}{3\mpl^2}} = \frac{\pi \fn{g^{1/2}_*}{T_\mathrm{d}} T_\mathrm{d}^2}{3\sqrt{10}\,\mpl} ,
\end{eqnarray}
therefore \eqss{kx}{ad}{Hd} give
\begin{eqnarray} \label{kd}
\frac{2\pi a_0}{k_\mathrm{d}} & = & \frac{ 6\sqrt{10} \fn{g_{*s}^{1/3}}{T_\mathrm{d}} \mpl }{ \fn{g_{*s}^{1/3}}{T_0} \fn{g_*^{1/2}}{T_\mathrm{d}} T_0 T_\mathrm{d} }
\\
& \simeq & \mathinner{0.4} \left[ \left( \frac{10^2 \fn{g_{*s}^2}{T_\mathrm{d}} }{\fn{g_*^3}{T_\mathrm{d}}} \right)^\frac{1}{6} \frac{\GeV}{T_\mathrm{d}} \right] \pc .
\end{eqnarray}

\subsection{The largest characteristic scale}

\eqss{ka}{kb}{kc} imply
\begin{eqnarray}
k_\mathrm{b} & < & k_\mathrm{a}, k_\mathrm{c} \, ,
\\
k_\mathrm{d} & < & k_\mathrm{c} \, ,
\end{eqnarray}
thus either $k_\mathrm{b}$ or $k_\mathrm{d}$ will be the largest scale.
\eqs{kb}{kc} give
\begin{equation} \label{kbkd}
k_\mathrm{b} = e^{-N_\mathrm{bc} + \frac{1}{2} N_\mathrm{cd}} k_\mathrm{d} \, ,
\end{equation}
therefore if
\begin{equation}
N_\mathrm{bc} > \frac{1}{2} N_\mathrm{cd} \, ,
\end{equation}
which we will assume, then $k_\mathrm{b}$ will be the largest physical scale and hence the one that can be most easily observed.
Using \eqss{Ncd}{kd}{kbkd},
\begin{eqnarray} \label{kblength}
\frac{2\pi a_0}{k_\mathrm{b}} & = & e^{N_\mathrm{bc}} \left[ \frac{ 720\sqrt{3}\,\pi \fn{g_{*s}}{T_\mathrm{d}} \mpl^3 }{ \fn{g_{*s}}{T_0} \fn{g_*}{T_\mathrm{d}} T_0^3 T_\mathrm{d} V_0^{1/2} } \right]^\frac{1}{3}
\\
& \simeq & \mathinner{6.7} \left[ \frac{e^{N_\mathrm{bc}}}{e^{20}} \left( \frac{\GeV}{T_\mathrm{d}}  \right)^\frac{1}{3} \left(  \frac{10^7\GeV}{V_0^{1/4}}  \right)^\frac{2}{3} \right] \kpc .
\end{eqnarray}

Modes with $k < k_\mathrm{b}$ remain outside the horizon throughout the thermal inflation eras, and so are not affected by thermal inflation, while those with $k > k_\mathrm{b}$ enter the horizon during moduli domination, allowing their evolution to be modified.
Hence, it is expected that there could be observable features of thermal inflation at $k \gtrsim k_\mathrm{b}$.

Modes with $k > k_\mathrm{a}$ enter the horizon before moduli domination and so probe that unknown era, and modes with $k > k_\mathrm{d}$ reenter the horizon during flaton domination and so will be twice modified.

In the next section, we study the evolution of the density perturbations for modes
\begin{equation}
k_\mathrm{b} \lesssim k \ll k_\mathrm{a}, k_\mathrm{d} \, .
\end{equation}

\section{Evolution of the density perturbations}
\label{sec:background}

During the moduli domination and thermal inflation eras, $t_\mathrm{a} < t < t_\mathrm{c}$, we have moduli matter (m), thermal radiation (r) and vacuum energy ($V_0$), with
\begin{eqnarray} \label{rho}
\rho & = & \rho_\mathrm{m} + \rho_\mathrm{r} + V_0 \, ,
\\ \label{p}
p & = & \frac{1}{3} \rho_\mathrm{r} - V_0 \, .
\end{eqnarray}
To describe the perturbations in moduli and radiation, we define the gauge invariant variables
\begin{eqnarray}
\Rm & \equiv & \mathcal{R} - \frac{H}{\dot\rho_\mathrm{m}} \delta\rho_\mathrm{m} \, ,
\\
\Rr & \equiv & \mathcal{R} - \frac{H}{\dot\rho_\mathrm{r}} \delta\rho_\mathrm{r} \, ,
\end{eqnarray}
where $\Rm$ is the curvature perturbation on constant moduli density hypersurfaces and $\Rr$ is the curvature perturbation on constant radiation density hypersurfaces.
\eq{rrhox} gives
\begin{eqnarray} \label{rrhom} &&
\ddot\Rm + H \left( 2 + \frac{\rho_\mathrm{m}}{\rho+p+\frac{2}{3}q^2} \right) \dot\Rm
- \frac{1}{3} q^2 \left( \frac{\rho_\mathrm{m}}{\rho+p+\frac{2}{3}q^2} \right) \Rm
\nonumber \\ && {}
= - \frac{\frac{4}{3} \rho_\mathrm{r}}{\rho+p+\frac{2}{3}q^2} \left( H \dot\Rr - \frac{1}{3} q^2 \Rr \right) ,
\\[1ex] \label{rrhor} &&
\ddot\Rr + H \left( 1 + \frac{\frac{8}{3} \rho_\mathrm{r}}{\rho+p+\frac{2}{3}q^2} \right) \dot\Rr
+ \frac{1}{3} q^2 \left( 1 - \frac{\frac{8}{3} \rho_\mathrm{r}}{\rho+p+\frac{2}{3}q^2} \right) \Rr
\nonumber \\ && {}
= - \frac{2\rho_\mathrm{m}}{\rho+p+\frac{2}{3}q^2} \left( H \dot\Rm - \frac{1}{3} q^2 \Rm \right) ,
\end{eqnarray}
where $q \equiv k/a$.

The physics at $t \sim t_\mathrm{a}$ is uncertain so we restrict ourselves to $t_\mathrm{a} \ll t < t_\mathrm{c}$, in which case
\begin{equation} \label{mdomr}
\rho_\mathrm{r} \ll \rho_\mathrm{m}
\end{equation}
and so \eqs{rrhom}{rrhor} reduce to
\begin{eqnarray} \label{rrhomev}
\ddot\Rm
+ H \left( 2 + \frac{\rho_\mathrm{m}}{\rho_\mathrm{m} + \frac{2}{3} q^2} \right) \dot\Rm
- \frac{1}{3} q^2 \left( \frac{\rho_\mathrm{m}}{\rho_\mathrm{m} + \frac{2}{3} q^2} \right) \Rm & = & 0 ,
\\ \label{rrhorev}
\ddot\Rr + H \dot\Rr + \frac{1}{3} q^2 \Rr 
& = & F ,
\end{eqnarray}
where
\begin{equation}
F = - 2 \left( \frac{\rho_\mathrm{m}}{\rho_\mathrm{m} + \frac{2}{3} q^2} \right) \left( H \dot\Rm - \frac{1}{3} q^2 \Rm \right) ,
\end{equation}
which have solution
\begin{eqnarray}
\fn{\Rm}{k,t} & = & \fn{A_\mathrm{m}}{k,t_\mathrm{i}} \left[ 1 + \frac{1}{3} \fn{H}{t} \int_{t_\mathrm{i}}^t \frac{\fn{q}{k,t'}^2 \d{t'}}{\fn{H}{t'}^2} \right]
+ \fn{B_\mathrm{m}}{k,t_\mathrm{i}} \frac{\fn{H}{t}}{\fn{H}{t_\mathrm{i}}} ,
\\ \label{Rrsol}
\fn{\Rr}{k,t}
& = & \fn{\Rr}{k,t_\mathrm{i}} \cos \int_{t_\mathrm{i}}^t \frac{\fn{q}{k,t'}\d{t'}}{\sqrt{3}}
+ \fn{\dot\Rr}{k,t_\mathrm{i}} \frac{\sqrt{3}}{\fn{q}{k,t_\mathrm{i}}} \sin \int_{t_\mathrm{i}}^t \frac{\fn{q}{k,t'}\d{t'}}{\sqrt{3}}
\nonumber \\ && {}
+ \int_{t_\mathrm{i}}^t \d{t'} \frac{\sqrt{3}}{\fn{q}{k,t'}} \sin \left( \int_{t'}^t \frac{\fn{q}{k,t''}\d{t''}}{\sqrt{3}} \right) \fn{F}{k,t'}
\end{eqnarray}
with
\begin{equation}
\fn{F}{k,t} = \fn{\rho_\mathrm{m}}{t} \left[ \frac{1}{3} \fn{A_\mathrm{m}}{k,t_\mathrm{i}} \fn{H}{t} \int_{t_\mathrm{i}}^t \frac{\fn{q}{k,t'}^2 \d{t'}}{\fn{H}{t'}^2}
+ \fn{B_\mathrm{m}}{k,t_\mathrm{i}} \frac{\fn{H}{t}}{\fn{H}{t_\mathrm{i}}} \right] ,
\end{equation}
where
\begin{eqnarray}
\fn{A_\mathrm{m}}{k,t_\mathrm{i}} & = & \frac{ \fn{\rho_\mathrm{m}}{t_\mathrm{i}} \fn{\Rm}{k,t_\mathrm{i}} + 2 \fn{H}{t_\mathrm{i}} \fn{\dot\Rm}{k,t_\mathrm{i}} }{ \fn{\rho_\mathrm{m}}{t_\mathrm{i}} + \frac{2}{3} \fn{q}{k,t_\mathrm{i}}^2 } ,
\\
\fn{B_\mathrm{m}}{k,t_\mathrm{i}} & = & \frac{ \frac{2}{3} \fn{q}{k,t_\mathrm{i}}^2 \fn{\Rm}{k,t_\mathrm{i}} - 2 \fn{H}{t_\mathrm{i}} \fn{\dot\Rm}{k,t_\mathrm{i}} }{ \fn{\rho_\mathrm{m}}{t_\mathrm{i}} + \frac{2}{3} \fn{q}{k,t_\mathrm{i}}^2 } .
\end{eqnarray}

Defining the curvature perturbation on constant density hypersurfaces and the entropy perturbation
\begin{eqnarray}
\Rt & \equiv & \frac{\dot\rho_\mathrm{m}}{\dot\rho} \Rm + \frac{\dot\rho_\mathrm{r}}{\dot\rho} \Rr
\simeq \Rm ,
\\
\Smr & \equiv & \Rr - \Rm ,
\end{eqnarray}
respectively, \eqs{rrhomev}{rrhorev} are equivalent to
\begin{eqnarray} \label{rrhomrev}
\ddot\Rt + H \left( 2 + \frac{\rho_\mathrm{m}}{\rho_\mathrm{m}+\frac{2}{3}q^2} \right) \dot\Rt
- \frac{1}{3} q^2 \left( \frac{\rho_\mathrm{m}}{\rho_\mathrm{m}+\frac{2}{3}q^2} \right) \Rt & = & 0 ,
\\ \label{smrev}
\ddot\Smr + H \dot\Smr + \frac{1}{3} q^2 \Smr & = & G ,
\end{eqnarray}
where
\begin{equation}
G = \frac{ \frac{2}{3} q^2 }{ \rho_\mathrm{m} + \frac{2}{3} q^2 } \left( H \dot\Rt - \frac{1}{3} q^2 \Rt \right) ,
\end{equation}
which have solution
\begin{eqnarray}
\fn{\Rt}{k,t} & = & \fn{A_\mathrm{m}}{k,t_\mathrm{i}} \left[ 1 + \frac{1}{3} \fn{H}{t} \int_{t_\mathrm{i}}^t \frac{\fn{q}{k,t'}^2 \d{t'}}{\fn{H}{t'}^2} \right] + \fn{B_\mathrm{m}}{k,t_\mathrm{i}} \frac{\fn{H}{t}}{\fn{H}{t_\mathrm{i}}} ,
\\
\fn{\Smr}{k,t}
& = & \left[ \fn{\Rr}{k,t_\mathrm{i}} - \fn{\Rm}{k,t_\mathrm{i}} \right] \cos \int_{t_\mathrm{i}}^t \frac{\fn{q}{k,t'}\d{t'}}{\sqrt{3}}
\nonumber \\ && {}
+ \left[ \fn{\dot\Rr}{k,t_\mathrm{i}} - \fn{\dot\Rm}{k,t_\mathrm{i}} \right] \frac{\sqrt{3}}{\fn{q}{k,t_\mathrm{i}}} \sin \int_{t_\mathrm{i}}^t \frac{\fn{q}{k,t'}\d{t'}}{\sqrt{3}}
\nonumber \\ && {}
+ \int_{t_\mathrm{i}}^t \d{t'} \frac{\sqrt{3}}{\fn{q}{k,t'}} \sin \left( \int_{t'}^t \frac{\fn{q}{k,t''}\d{t''}}{\sqrt{3}} \right) \fn{G}{k,t'}
\end{eqnarray}
with
\begin{equation}
\fn{G}{k,t} = - \frac{1}{3} \fn{q}{k,t}^2 \left[ \frac{1}{3} \fn{A_\mathrm{m}}{k,t_\mathrm{i}} \fn{H}{t} \int_{t_\mathrm{i}}^t \frac{\fn{q}{k,t'}^2 \d{t'}}{\fn{H}{t'}^2}
+ \fn{B_\mathrm{m}}{k,t_\mathrm{i}} \frac{\fn{H}{t}}{\fn{H}{t_\mathrm{i}}} \right] .
\end{equation}

Thermal inflation ends when the temperature of the radiation drops to a critical temperature, given by \eq{Tc}, triggering a first order phase transition converting the vacuum energy $V_0$ into flaton matter.
The transition hypersurface matches to constant radiation density hypersurfaces before the transition and constant density hypersurfaces after the transition.
Thus the late time perturbation is given by identifying
\begin{equation}
\fn{\Rt}{k,t_\mathrm{c}^+} = \fn{\Rr}{k,t_\mathrm{c}^-} .
\end{equation}
Using \eq{Rrsol} and taking $t_\mathrm{i} < t_\mathrm{c} < t_\mathrm{f}$ gives
\begin{eqnarray} \label{Rtq}
\fn{\Rt}{k,t_\mathrm{f}}
& = & \fn{\Rr}{k,t_\mathrm{i}} \cos \int_{t_\mathrm{i}}^{t_\mathrm{f}} \frac{\fn{q}{k,t}\d{t}}{\sqrt{3}}
+ \frac{ \sqrt{3} \fn{\dot\Rr}{k,t_\mathrm{i}} }{ \fn{q}{k,t_\mathrm{i}} } \sin \int_{t_\mathrm{i}}^{t_\mathrm{f}} \frac{\fn{q}{k,t}\d{t}}{\sqrt{3}}
\nonumber \\ && {}
+ \frac{1}{3} \fn{A_\mathrm{m}}{k,t_\mathrm{i}} \int_{t_\mathrm{i}}^{t_\mathrm{f}} \d{t} \frac{ \sqrt{3} \fn{H}{t} \fn{\rho_\mathrm{m}}{t} }{ \fn{q}{k,t} } \int_{t_\mathrm{i}}^t \frac{\fn{q}{k,t'}^2 \d{t'}}{\fn{H}{t'}^2} \sin \int_t^{t_\mathrm{f}} \frac{\fn{q}{k,t''}\d{t''}}{\sqrt{3}}
\nonumber \\ && {}
+ \frac{ \fn{B_\mathrm{m}}{k,t_\mathrm{i}} }{ \fn{H}{t_\mathrm{i}} } \int_{t_\mathrm{i}}^{t_\mathrm{f}} \d{t} \frac{ \sqrt{3} \fn{H}{t} \fn{\rho_\mathrm{m}}{t} }{ \fn{q}{k,t} } \sin \int_t^{t_\mathrm{f}} \frac{\fn{q}{k,t'}\d{t'}}{\sqrt{3}} .
\end{eqnarray}

The scale
\begin{equation}
k_\mathrm{b} \equiv \fn{a}{t_\mathrm{b}} \fn{H}{t_\mathrm{b}} ,
\end{equation}
discussed in Section~\ref{sec:history}, is of central importance in this paper, and we precisely define $t_\mathrm{b}$, the boundary between moduli domination and thermal inflation, by
\begin{equation} \label{ddota}
\ddot{a}(t_\mathrm{b}) \equiv 0
\end{equation}
or equivalently, using \eq{mdomr},
\begin{equation} \label{rhomb}
\fn{\rho_\mathrm{m}}{t_\mathrm{b}} \simeq 2 V_0 \, .
\end{equation}
Defining
\begin{eqnarray} \label{alpha}
\alpha & \equiv & \frac{a}{a_\mathrm{b}} ,
\\ \label{kappa}
\kappa & \equiv & \frac{k}{k_\mathrm{b}}
\end{eqnarray}
and using \eqs{mdomr}{rhomb} gives
\begin{equation} \label{Halpha}
3  H^2 \simeq V_0 + \rho_\mathrm{m} = V_0 \left( 1 + \frac{2}{\alpha^3} \right)
\end{equation}
and \eq{Rtq} becomes
\begin{eqnarray}
\fn{\Rt}{\kappa,\alpha_\mathrm{f}}
& = & \fn{\Rr}{\kappa,\alpha_\mathrm{i}} \cos \left[ \kappa \int_{\alpha_\mathrm{i}}^{\alpha_\mathrm{f}} \frac{d\alpha}{\sqrt{\alpha(2+\alpha^3)}} \right]
\nonumber \\ && {}
+ \frac{1}{\kappa} \left( \frac{2+\alpha_\mathrm{i}^3}{\alpha_\mathrm{i}} \right)^\frac{1}{2} \fn{\frac{d\Rr}{d\ln\alpha}}{\kappa,\alpha_\mathrm{i}}
\sin \left[ \kappa \int_{\alpha_\mathrm{i}}^{\alpha_\mathrm{f}} \frac{d\alpha}{\sqrt{\alpha(2+\alpha^3)}} \right] \nonumber \\ && {}
+ 6 \kappa \fn{A_\mathrm{m}}{\kappa,\alpha_\mathrm{i}} \int_{\alpha_\mathrm{i}}^{\alpha_\mathrm{f}} \frac{d\alpha}{\alpha^3} \int_{\alpha_\mathrm{i}}^\alpha d\beta
\left( \frac{\beta}{2+\beta^3} \right)^\frac{3}{2}
\sin \left[ \kappa \int_\alpha^{\alpha_\mathrm{f}} \frac{d\gamma}{\sqrt{\gamma(2+\gamma^3)}} \right]
\nonumber \\ && {}
+ \frac{6}{\kappa} \left( \frac{\alpha_\mathrm{i}^3}{2+\alpha_\mathrm{i}^3} \right)^\frac{1}{2} \fn{B_\mathrm{m}}{\kappa,\alpha_\mathrm{i}} \int_{\alpha_\mathrm{i}}^{\alpha_\mathrm{f}} \frac{d\alpha}{\alpha^3} \sin \left[ \kappa \int_\alpha^{\alpha_\mathrm{f}} \frac{d\beta}{\sqrt{\beta(2+\beta^3)}} \right] ,
\end{eqnarray}
where
\begin{eqnarray}
\fn{A_\mathrm{m}}{\kappa,\alpha_\mathrm{i}} & = & \frac{1}{1+\frac{1}{3}\kappa^2\alpha_\mathrm{i}} \left[ \fn{\Rm}{\kappa,\alpha_\mathrm{i}} + \frac{1}{3} \left( 2 + \alpha_\mathrm{i}^3 \right) \fn{\frac{d\Rm}{d\ln\alpha}}{\kappa,\alpha_\mathrm{i}} \right] ,
\\
\fn{B_\mathrm{m}}{\kappa,\alpha_\mathrm{i}} & = & \frac{1}{1+\frac{1}{3}\kappa^2\alpha_\mathrm{i}} \left[ \frac{1}{3} \kappa^2 \alpha_\mathrm{i} \fn{\Rm}{\kappa,\alpha_\mathrm{i}} - \frac{1}{3} \left( 2 + \alpha_\mathrm{i}^3 \right) \fn{\frac{d\Rm}{d\ln\alpha}}{\kappa,\alpha_\mathrm{i}} \right] .
\end{eqnarray}
Taking $t_\mathrm{i} \ll t_\mathrm{b}$ and
$q_\mathrm{i} \ll H_\mathrm{i}$, i.e.\ $\alpha_\mathrm{i} \ll 1$ and $\kappa^2 \alpha_\mathrm{i} \ll 1$, gives
\begin{eqnarray} \label{Rsollate}
\fn{\Rt}{\kappa,\alpha_\mathrm{f}}
& = & \fn{\Rr}{\kappa,\alpha_\mathrm{i}} \cos \left[ \kappa \int_{\alpha_\mathrm{i}}^{\alpha_\mathrm{f}} \frac{d\alpha}{\sqrt{\alpha(2+\alpha^3)}} \right]
\nonumber \\ && {}
+ \frac{1}{\kappa} \sqrt{\frac{2}{\alpha_\mathrm{i}}} \fn{\frac{d\Rr}{d\ln\alpha}}{\kappa,\alpha_\mathrm{i}} \sin \left[ \kappa \int_{\alpha_\mathrm{i}}^{\alpha_\mathrm{f}} \frac{d\alpha}{\sqrt{\alpha(2+\alpha^3)}} \right] \nonumber \\ && {}
+ 6 \kappa \fn{A_\mathrm{m}}{\kappa,\alpha_\mathrm{i}} \int_{\alpha_\mathrm{i}}^{\alpha_\mathrm{f}} \frac{d\alpha}{\alpha^3} \int_{\alpha_\mathrm{i}}^\alpha d\beta
\left( \frac{\beta}{2+\beta^3} \right)^\frac{3}{2}
\sin \left[ \kappa \int_\alpha^{\alpha_\mathrm{f}} \frac{d\gamma}{\sqrt{\gamma(2+\gamma^3)}} \right]
\nonumber \\ && {}
+ \frac{3}{\kappa} \sqrt{2\alpha_\mathrm{i}^3} \fn{B_\mathrm{m}}{\kappa,\alpha_\mathrm{i}}  \int_{\alpha_\mathrm{i}}^{\alpha_\mathrm{f}} \frac{d\alpha}{\alpha^3} \sin \left[ \kappa \int_\alpha^{\alpha_\mathrm{f}} \frac{d\beta}{\sqrt{\beta(2+\beta^3)}} \right] ,
\end{eqnarray}
where
\begin{eqnarray}
\fn{A_\mathrm{m}}{\kappa,\alpha_\mathrm{i}}
& = & \fn{\Rm}{\kappa,\alpha_\mathrm{i}} + \frac{2}{3} \fn{\frac{d\Rm}{d\ln\alpha}}{\kappa,\alpha_\mathrm{i}} ,
\\
\fn{B_\mathrm{m}}{\kappa,\alpha_\mathrm{i}}
& = & \frac{1}{3} \kappa^2 \alpha_\mathrm{i} \fn{\Rm}{\kappa,\alpha_\mathrm{i}} - \frac{2}{3} \fn{\frac{d\Rm}{d\ln\alpha}}{\kappa,\alpha_\mathrm{i}} .
\end{eqnarray}

\subsection{Thermal inflation transfer function}

\begin{figure}
\centering
\includegraphics[width=0.8\textwidth]{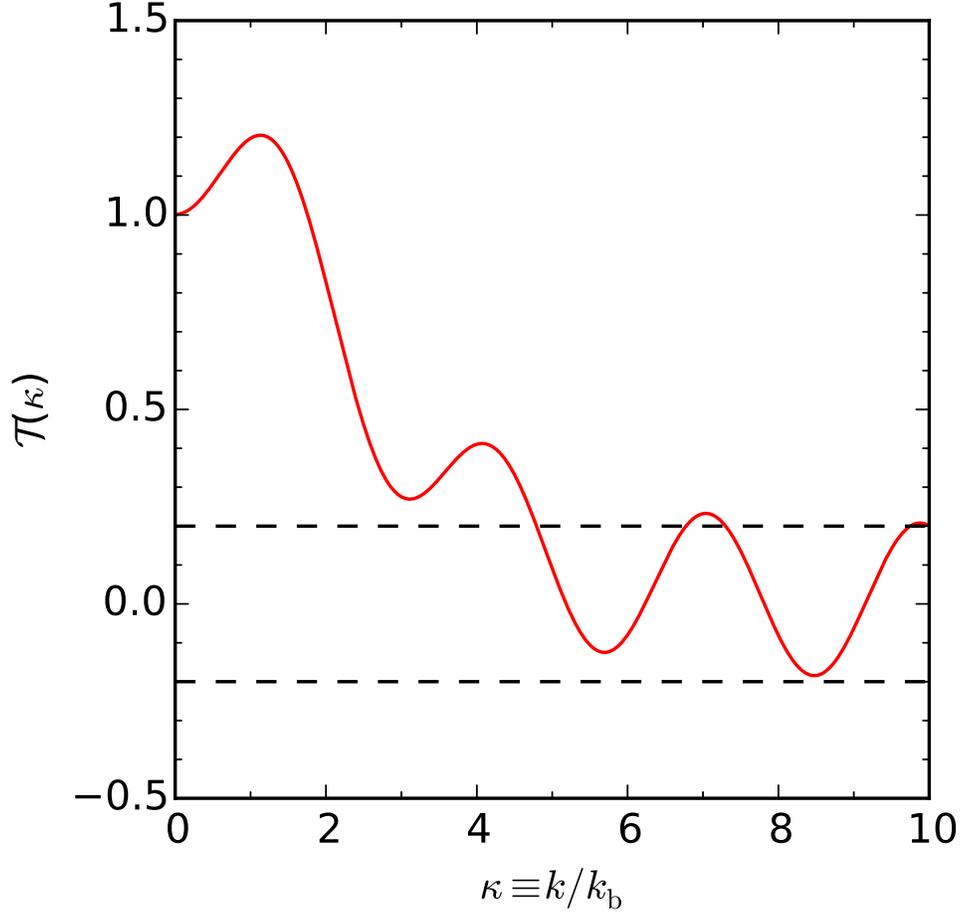}
\caption{ \label{fig:Rdeltarhor}
The thermal inflation transfer function $\fn{\mathcal{T}}{\kappa}$, where
$\kappa \equiv k/k_\mathrm{b}$ and the characteristic scale $k_\mathrm{b}$ is defined by \eq{ddota} and estimated in \eq{kblength}.
The first peak is at $(\kappa,\mathcal{T}) \simeq (1.13,1.21)$ and the first dip is at $(3.11,0.269)$.
The analytic form is given in \eq{latetime} and the asymptotic behaviours are given in \eqs{kappazero}{kappainfty}.
}
\end{figure}

If we take an adiabatic initial condition
\begin{equation}
\fn{\Smr}{k,t_\mathrm{i}} = \fn{\dot\Smr}{k,t_\mathrm{i}} = 0
\end{equation}
so that
\begin{eqnarray}
\fn{\Rm}{k,t_\mathrm{i}} = \fn{\Rr}{k,t_\mathrm{i}}
& = & \fn{\Rt}{k,t_\mathrm{i}} ,
\\
\fn{\dot\Rm}{k,t_\mathrm{i}} = \fn{\dot\Rr}{k,t_\mathrm{i}}
& = & \fn{\dot\Rt}{k,t_\mathrm{i}} 
\end{eqnarray}
neglect the decaying mode, $\fn{B_\mathrm{m}}{k,t_\mathrm{i}} = 0$, and take $t_\mathrm{i} \to 0$ and $t_\mathrm{f} \to \infty$, the effect of the thermal inflation era on the curvature perturbation can be expressed as the transfer function
\begin{eqnarray}
\fn{\mathcal{T}}{\kappa} \equiv \frac{\fn{\Rt}{\kappa,\infty}}{\fn{\Rt}{\kappa,0}} ,
\end{eqnarray}
which, from \eq{Rsollate}, is given by
\begin{eqnarray} \label{latetime}
\fn{\mathcal{T}}{\kappa} & = & \cos \left[ \kappa \int_0^\infty \frac{d\alpha}{\sqrt{\alpha(2+\alpha^3)}} \right] \nonumber \\ && {}
+ 6 \kappa \int_0^\infty \frac{d\gamma}{\gamma^3} \int_0^\gamma d\beta \left( \frac{\beta}{2+\beta^3}\right)^\frac{3}{2}
\sin \left[ \kappa \int_\gamma^\infty \frac{d\alpha}{\sqrt{\alpha(2+\alpha^3)}} \right]
\end{eqnarray}
and is plotted in Figure~\ref{fig:Rdeltarhor}.
On large scales, which remain outside the horizon, the transfer function asymptotes to one
\begin{equation} \label{kappazero}
\fn{\mathcal{T}}{\kappa} \stackrel{\kappa \to 0}{\longrightarrow}
1 + \nu_0 \kappa^2 + \fn{\mathcal{O}}{\kappa^4} ,
\end{equation}
where, using \eq{nu0calc},
\begin{equation}
\nu_0 = \int_0^\infty \d\alpha \left( \frac{\alpha}{2+\alpha^3} \right)^\frac{3}{2}
= \frac{2^{7/3}\pi^{3/2}}{3^{3/2} \fn{\Gamma}{\frac{1}{6}} \fn{\Gamma}{\frac{1}{3}}}
\simeq 0.3622 .
\end{equation}
On small scales, which enter well into the horizon during moduli domination and thermal inflation, the transfer function oscillates due to the oscillation of the radiation perturbation inside the horizon
\begin{equation} \label{kappainfty}
\fn{\mathcal{T}}{\kappa} \stackrel{\kappa \to \infty}{\longrightarrow} - \frac{1}{5} \cos(\nu_1 \kappa) + \fn{o}{\kappa^{-n}} ,
\end{equation}
where we have used \eq{nu1calc} and
\begin{equation}
\nu_1 = \int_0^\infty \frac{\d\alpha}{\sqrt{\alpha(2+\alpha^3)}}
= \frac{\fn{\Gamma}{\frac{1}{6}} \fn{\Gamma}{\frac{1}{3}}}{2^{1/3}3\sqrt{\pi}}
\simeq 2.2258 .
\end{equation}

\begin{figure}
\centering
\includegraphics[width=0.8\textwidth]{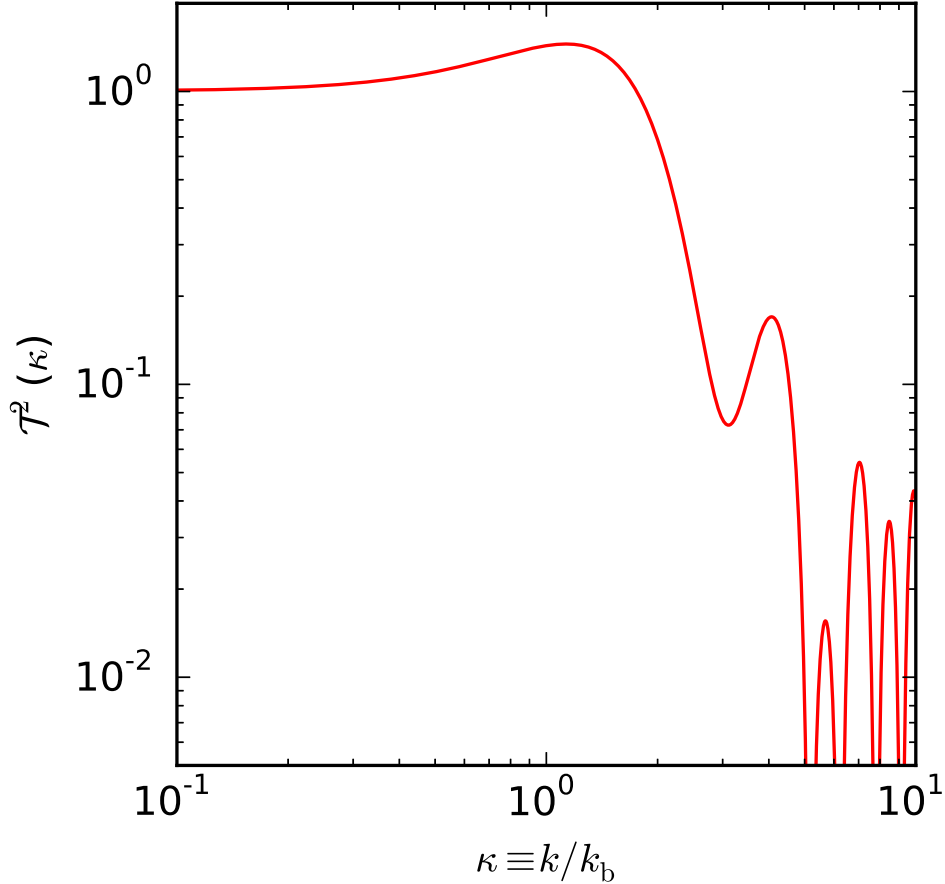}
\caption{ \label{fig:R2deltarhor}
The thermal inflation transfer function $\fn{\mathcal{T}^2}{\kappa}$.
}
\end{figure}

The power spectrum after thermal inflation is
\begin{equation}
\fn{P}{k} = \fn{\mathcal{T}^2}{k/k_\mathrm{b}} \times \fn{P_\mathrm{pri}}{k} ,
\end{equation}
where $\fn{P_\mathrm{pri}}{k}$ is the power spectrum of the primordial inflation, see Figure~\ref{fig:R2deltarhor}.
Hence, the power spectrum is enhanced by a factor $1.46$ at the first peak and, from Eq.~(\ref{kappainfty}), suppressed by a factor $50$ on small scales.

\section{Discussion}
\label{sec:discussion}

In this paper, we presented the effect of thermal inflation on the linear evolution of small scale density perturbations.
By introducing thermal inflation, the post-inflationary history is changed to include extra eras of moduli matter domination, thermal inflation and flaton matter domination, followed by the usual radiation domination, as in Figure~\ref{fig:4scales}.
Modes with $k > k_\mathrm{b}$ enter the horizon during moduli domination before exiting again during thermal inflation, and hence are modified relative to the modes with $k < k_\mathrm{b}$ which remain outside the horizon.
At the end of thermal inflation, the radiation perturbation is converted into the final adiabatic (curvature) perturbation.
The net effect of this for an adiabatic primordial perturbation is to suppress the perturbations by a factor $\sim 50$ on scales $k \gg k_\mathrm{b}$, see Figures~\ref{fig:Rdeltarhor} and~\ref{fig:R2deltarhor}, and Eq.~(\ref{kappainfty}).

A multicomponent, i.e.\ adiabatic (curvature) plus entropy (isocurvature), primordial perturbation  leads to a more complicated, initial condition dependent, result given in \eq{Rsollate}.
Even for modes with $k < k_\mathrm{b}$, which remain outside the horizon, the case with thermal inflation can lead to different results compared to without thermal inflation as it is the subdominant radiation perturbation, rather than the dominant moduli matter perturbation, that is converted to the final adiabatic perturbation.

The scale $k_\mathrm{b}$ is theoretically estimated in \eq{kblength}, but the uncertainties are large, with anything from megaparsec to subparsec scales being reasonable, and no robust upper or lower bound.
Current and future observations could provide stronger constraints, or an observational signature of thermal inflation.
For example, the lack of small scale suppression in the primordial power spectrum reconstructed from the recent Planck results \cite{Ade:2015lrj} suggests $k_\mathrm{b} \gtrsim 1 \Mpc^{-1}$.

The suppression of the power spectrum would reduce the number density of dark matter clumps of mass
\begin{equation}
M \lesssim \frac{4\pi}{3} \fn{\rho_\mathrm{m}}{t_0} \left( \frac{2\pi a_0}{3 k_\mathrm{b}} \right)^3
\sim 10^{11} M_{\odot} \left( \frac{k_\mathrm{b}}{\Mpc^{-1}} \right)^{-3} ,
\end{equation}
where the scale $3k_\mathrm{b}$ is estimated from Figure~\ref{fig:R2deltarhor}.
Depending on the value of $k_\mathrm{b}$, this might be able to be seen from 21cm hydrogen line observations such as the Square Kilometre Array (SKA), see for example \cite{Furlanetto:2009qk}, or gamma-ray observations of WIMP-annihilation in, or gravitational microlensing of, ultracompact minihalos, see for example \cite{Bringmann:2011ut}.

Silk damping might be relevant when radiation perturbations enter the horizon during either $t_\mathrm{a} < t < t_\mathrm{c}$ or $t > t_\mathrm{d}$.
For $t_\mathrm{a} < t < t_\mathrm{c}$, we estimate the Silk damping scale to be much smaller than our characteristic scale $k_\mathrm{b}$ \cite{Jeong:2014gna}.
For $t > t_\mathrm{d}$, modes near $k_\mathrm{b}$ can be affected by Silk damping.
For $50\Mpc^{-1} \lesssim  k_\mathrm{b} \lesssim 10^4 \Mpc^{-1}$, the dissipation of modes with $k \sim k_\mathrm{b}$ would generate a unique pattern of CMB distortions, different from \cite{Chluba:2012we,Khatri:2011aj,Chluba:2012gq,Khatri:2013dha}, due to the suppression and oscillations of the thermal inflation transfer function.
Future observations such as the Primordial Inflation Explorer (PIXIE) \cite{Kogut:2011xw} or the Polarized Radiation Imaging and Spectroscopy Mission (PRISM) \cite{Andre:2013afa, Andre:2013nfa} could be used to probe these distortions \cite{Chluba:2012we}.
For $10^4\Mpc^{-1} \lesssim k_\mathrm{b} \lesssim 10^5\Mpc^{-1}$, the suppression of the power spectrum could diminish any effect of the release of energy by Silk damping on Big Bang nucleosynthesis \cite{Jeong:2014gna}.

\begin{acknowledgments}

The authors thank Raghavan Rangarajan, Donghui Jeong, Ido Ben-Dayan, Wan-il Park, Richard Easther, Chang Sub Shin, Ki-Young Choi, Kyungjin Ahn, Bayram Tekin, Dong-han Yeom and Kyujin Kwak for useful discussions at various stages of this work.
HZ thanks KAIST, APCTP and METU for their hospitality.
This work was supported by the Basic Science Research Program through the National Research Foundation of
Korea (NRF) funded by the Ministry of Education, Science and Technology (N01110095 and N01130488).

\end{acknowledgments}

\appendix

\section{Review of perturbations in a multicomponent system} \label{sec:general}

The scalar parts of the perturbed metric and energy momentum tensor are \cite{Kodama:1985bj}
\begin{equation}
\tilde{ds}^2 = (1+2A) \d{t}^2 - 2 B_{,i} \d{t} \d{x^i} - \left[ (1+2\mathcal{R}) \fn{a^2}{t} \delta_{ij} + 2 C_{,ij} \right] \d{x^i} \d{x^j}
\end{equation}
and
\begin{equation}
\tilde{T}_{\mu\nu} = \tilde{\rho} \tilde{u}_\mu \tilde{u}_\nu - \tilde{p} \left( \tilde{g}_{\mu\nu} - \tilde{u}_\mu \tilde{u}_\nu \right) + \tilde\pi_{\mu\nu} \, ,
\end{equation}
where $\tilde{u}^0 = 1/a$ and
\begin{equation}
\begin{array}{RCLCRCL}
\tilde\rho & = & \rho + \delta\rho
& \comma &
\tilde{u}^i & = & \frac{1}{a^2} \partial_i v ,
\\[1ex]
\tilde{p} & = & p + \delta p
& \comma &
\tilde\pi_{\mu\nu} & = & \partial_i \partial_j \pi .
\end{array}
\end{equation}
The Einstein equation gives
\begin{eqnarray} \label{einstein1}
3 H \left( \dot\mathcal{R} - H A \right) + q^2 \left[ \mathcal{R} - H \left( \dot{C} - 2 H C - B \right) \right]
& = & \frac{1}{2} \delta\rho ,
\\ \label{einstein2}
\dot\mathcal{R} - H A & = & - \dot{H} \left( v + B \right) ,
\\ \label{einstein3}
\frac{d}{dt} \left( \dot{C} - 2 H C - B \right) + H \left( \dot{C} - 2 H C - B \right) - \mathcal{R} - A
& = & \pi ,
\\ \label{einstein4}
\frac{d}{dt} \left( \dot\mathcal{R} - H A \right) + 3 H \left( \dot\mathcal{R} - H A \right) - \dot{H} A
& = & - \frac{1}{2} \delta p + \frac{1}{3} q^2 \pi ,
\end{eqnarray}
where $q \equiv k/a$.
Taking the derivatives of \eqs{einstein1}{einstein2} gives
\begin{equation} \label{div1}
\dot{\delta\rho} + 3 H \left( \delta\rho + \delta p \right) + 2 q^2 \dot{H} \left( v + B \right) - 2 \dot{H} \left[ 3 \dot\mathcal{R} - q^2 \left( \dot{C} - 2 H C - B \right) \right] = 0 ,
\end{equation}
\begin{equation} \label{div2}
\frac{d}{dt} \left[ \dot{H} \left( v + B \right) \right] + 3 H \dot{H} \left( v + B \right) - \frac{1}{2} \delta p
+ \frac{1}{3} q^2 \pi + \dot{H} A = 0
\end{equation}
corresponding to $\nabla \cdot T = 0$.
Decomposing into components
\begin{equation}
\begin{array}{RCLCRCL}
\delta\rho  & = & \sum_X \delta \rho_X
& \comma &
\left( \rho + p \right) v & = & \sum_X \left( \rho_X + p_X \right) v_X ,
\\
\delta p & = & \sum_X \delta p_X
& \comma &
\pi & = & \sum_X \pi_X
\end{array}
\end{equation}
gives
\begin{equation}
\sum_X \left\{ \dot{\delta\rho}_X + 3 H \left( \delta\rho_X + \delta p_X \right) - q^2 \left( \rho_X + p_X \right) \left( v_X + B \right) + \left( \rho_X + p_X \right) \left[ 3 \dot\mathcal{R} - q^2 \left( \dot{C} - 2 H C - B \right) \right] \right\} = 0 ,
\end{equation}
\begin{equation}
\sum_X \left\{ \frac{d}{dt} \left[ \left( \rho_X + p_X \right) \left( v_X + B \right) \right] + 3 H \left( \rho_X + p_X \right) \left( v_X + B \right) + \delta p_X - \frac{2}{3} q^2 \pi_X + \left( \rho_X + p_X \right) A \right\} = 0
\end{equation}
corresponding to $\sum_X \nabla \cdot T_X = 0$.

For components which couple only gravitationally, we have $\nabla \cdot T_{X} = 0$ separately
\begin{equation} \label{singlediv1}
\dot{\delta\rho}_X + 3 H \left( \delta\rho_X + \delta p_X \right) - q^2 \left( \rho_X + p_X \right) \left( v_X + B \right)
+ \left( \rho_X + p_X \right) \left[ 3 \dot\mathcal{R} -q^2 \left( \dot{C} - 2 H C - B \right) \right] = 0 ,
\end{equation}
\begin{equation} \label{singlediv2}
\frac{d}{dt} \left[ \left( \rho_X + p_X \right) \left( v_X + B \right) \right] + 3 H \left( \rho_X + p_X \right) \left( v_X + B \right) + \delta p_X - \frac{2}{3} q^2 \pi_X + \left( \rho_X + p_X \right) A = 0 .
\end{equation}
For
\begin{eqnarray}
\delta p_X & = & \frac{\dot{p}_X}{\dot\rho_X} \delta\rho_X \, ,
\\
\pi_X & = & 0 ,
\end{eqnarray}
combining \eqs{singlediv1}{singlediv2} and using Eqs.~(\ref{einstein1}) to~(\ref{einstein4}) gives
\begin{eqnarray} \label{single0}
\lefteqn{
\frac{d^2}{dt^2}\left( \mathcal{R} - \frac{H}{\dot\rho_X} \delta\rho_X \right)
+ H \left( 2 - 3 \frac{\dot{p}_X}{\dot\rho_X} \right)
\frac{d}{dt}\left( \mathcal{R} - \frac{H}{\dot\rho_X} \delta\rho_X \right)
+ q^2 \frac{\dot{p}_X}{\dot\rho_X}
\left( \mathcal{R} - \frac{H}{\dot\rho_X} \delta\rho_X \right)
} \nonumber \\
& = & \frac{1}{3} \left( \frac{1 + 3 \frac{\dot{p}_X}{\dot\rho_X}}{\rho+p+\frac{2}{3}q^2} \right) \left[
\dot\rho \frac{d}{dt} \left( \mathcal{R} - \frac{H}{\dot\rho} \delta\rho \right) + q^2 \left( \rho + p \right) \left( \mathcal{R} - \frac{H}{\dot\rho} \delta\rho \right) + 3 H^2 \dot{p} \left( \frac{\delta\rho}{\dot\rho} - \frac{\delta p}{\dot{p}} \right)
\right]
\nonumber \\
\end{eqnarray}
with
\begin{eqnarray}
\mathcal{R} - \frac{H}{\dot\rho} \delta\rho
& = & \sum_X \frac{\dot\rho_X}{\dot\rho} \left( \mathcal{R} - \frac{H}{\dot\rho_X} \delta\rho_X \right) ,
\\
H \left( \frac{\delta\rho}{\dot\rho} - \frac{\delta p}{\dot{p}} \right)
& = & - \sum_X \left( \frac{\dot\rho_X}{\dot\rho} - \frac{\dot{p}_X}{\dot{p}} \right) \left( \mathcal{R} - \frac{H}{\dot\rho_X} \delta\rho_X \right) .
\end{eqnarray}
For two components, $X$ and $Y$, \eq{single0} gives
\begin{eqnarray}
\lefteqn{
\ddot\RX
+ H \left[ 3 - \left( 1 + 3 \frac{\dot{p}_X}{\dot\rho_X} \right) \left( \frac{\rho_Y+p_Y+\frac{2}{3}q^2}{\rho+p+\frac{2}{3}q^2} \right) \right] \dot\RX
} \nonumber \\ \lefteqn{ {}
- \frac{1}{3} q^2 \left[ 1 - \left( 1 + 3 \frac{\dot{p}_X}{\dot\rho_X} \right) \left( \frac{\rho_Y+p_Y+\frac{2}{3}q^2}{\rho+p+\frac{2}{3}q^2} \right) \right] \RX
} \nonumber \\ \label{rrhox}
& = & - \left( 1 + 3 \frac{\dot{p}_X}{\dot\rho_X} \right) \left( \frac{\rho_Y + p_Y}{\rho+p+\frac{2}{3}q^2} \right) \left( H \dot\RY - \frac{1}{3} q^2 \RY \right) ,
\end{eqnarray}
where
\begin{equation}
\RX \equiv \mathcal{R} - \frac{H}{\dot\rho_X} \delta\rho_X .
\end{equation}

\section{Mathematical formulae}
\label{math}

Integrating by parts:
\begin{eqnarray} \label{nu0calc}
\lefteqn{
6 \int_0^\infty \frac{\d\gamma}{\gamma^3}
\int_0^\gamma \d\beta \left( \frac{\beta}{2+\beta^3} \right)^\frac{3}{2} \int_\gamma^\infty
\frac{\d\alpha}{\sqrt{\alpha(2+\alpha^3)}}
} \nonumber \\
& = & 3 \int_0^\infty \frac{\d\gamma}{\gamma^{1/2}\left(2+\gamma^3\right)^{3/2}}
\int_\gamma^\infty \frac{\d\alpha}{\sqrt{\alpha(2+\alpha^3)}}
- 3 \int_0^\infty \frac{\d\gamma}{\gamma^{5/2}\left(2+\gamma^3\right)^{1/2}} \int_0^\gamma \d\beta \left( \frac{\beta}{2+\beta^3} \right)^\frac{3}{2}
\nonumber \\
& = & \frac{1}{2} \left[ \int_0^\infty \frac{\d\alpha}{\sqrt{\alpha(2+\alpha^3)}} \right]^2 + \int_0^\infty \d\alpha \left( \frac{\alpha}{2+\alpha^3} \right)^\frac{3}{2} ,
\end{eqnarray}
where we have used
\begin{equation}
\int \frac{ 3 \gamma^{1/2} \d\gamma }{ \left(2+\gamma^3\right)^{3/2} } = \frac{ \gamma^{3/2} }{ \left(2+\gamma^3\right)^{1/2} }
\end{equation}
and
\begin{equation}
\int \frac{ 3 \d\gamma }{ \gamma^{5/2} \left(2+\gamma^3\right)^{1/2} }
= - \frac{ \left(2+\gamma^3\right)^{1/2} }{ \gamma^{3/2} } .
\end{equation}

Integrating by parts:
\begin{eqnarray} \label{nu1calc}
\lefteqn{
6 \kappa \int_0^\infty \frac{d\gamma}{\gamma^3} \sin \left[ \kappa \int_\gamma^\infty \frac{d\alpha}{\sqrt{\alpha(2+\alpha^3)}} \right]
\int_0^\gamma d\beta \left( \frac{\beta}{2+\beta^3}\right)^\frac{3}{2}
} \nonumber \\
& = & - \frac{6}{5} \cos \left[ \kappa \int_0^\infty \frac{d\alpha}{\sqrt{\alpha(2+\alpha^3)}} \right]
- 9 \int_0^\infty \frac{d\gamma}{\sqrt{\gamma(2+\gamma^3)}}
\cos \left[ \kappa \int_\gamma^\infty \frac{d\alpha}{\sqrt{\alpha(2+\alpha^3)}} \right]
\nonumber \\ && {} \times
\left[ \int_0^\gamma \d\beta \left( \frac{\beta}{2+\beta^3}\right)^\frac{3}{2} - 3 \left( \frac{2+\gamma^3}{\gamma^3} \right) \int_0^\gamma \d\beta \frac{\beta^{9/2}}{(2+\beta^3)^{5/2}} \right]
\nonumber \\
& = & - \frac{6}{5} \cos \left[ \kappa \int_0^\infty \frac{d\alpha}{\sqrt{\alpha(2+\alpha^3)}} \right]
\nonumber \\ && {}
+ \frac{18}{\kappa} \int_0^\infty d\gamma
\sin \left[ \kappa \int_\gamma^\infty \frac{d\alpha}{\sqrt{\alpha(2+\alpha^3)}} \right]
\left[ \left( \frac{\gamma}{2+\gamma^3}\right)^\frac{3}{2}
- \frac{9}{\gamma^4} \int_0^\gamma \d\beta \frac{\beta^{9/2}}{(2+\beta^3)^{5/2}} \right] .
\end{eqnarray}

\end{document}